\documentclass{article}
\usepackage{amsmath,amssymb}
\usepackage{graphicx}
\usepackage[utf8]{inputenc}
\usepackage{epstopdf}
\usepackage{url}
\usepackage{tikz,pgf}
\usepackage{pgf-pie}
\usepackage{subfig}
\usetikzlibrary{patterns}
\usepackage{arydshln}
\usepackage{todonotes}

\newcommand{\tod}[1]{}

\DeclareMathOperator{\cost}{cost}

\begin{document}
\sloppy

\title{LiveRank: How to Refresh Old Datasets}

\author{
	The Dang Huynh\\
	Alcatel-Lucent Bell Labs France
	\and 
	Fabien Mathieu\\
	Alcatel-Lucent Bell Labs France
	\and
	Laurent Viennot\\
	Inria -- Univ. Paris Diderot
}

\maketitle
\begin{abstract}
This paper considers the problem of refreshing a dataset. More precisely, given a collection of nodes gathered at some time (Web pages, users from an online social network) along with some structure (hyperlinks, social relationships), we want to identify a significant fraction of the nodes that still exist at present time. The liveness of an old node can be tested through an online query at present time. We call LiveRank a ranking of the old pages so that active nodes are more likely to appear first. The quality of a LiveRank is measured by the number of queries necessary to identify a given fraction of the active nodes when using the LiveRank order. We study different scenarios from a static setting where the LiveRank is computed before any query is made, to dynamic settings where the LiveRank can be updated as queries are processed. Our results show that building on the PageRank can lead to efficient LiveRanks, for Web graphs as well as for online social networks.
\end{abstract}

\section{Introduction}

\label{subsec:motivation}

One of the main challenges for large networks data mining is to deal with the high dynamics of huge datasets: not only are these datasets difficult to gather, but they tend to become obsolete very quickly.

In this paper, we are interested in the evolution at large time scale of any large corpus available online.
Our primary focus will be the Web, but our approach encompasses any online data with similar linkage enabling crawling, like P2P networks or online social networks.
We thus focus on batch crawling, where starting from a completely out-dated snapshot of a large graph like the Web, we want to identify a significant fraction of the nodes that are still alive now. The interest is twofold. 

First, many old snapshot of large graphs are available today. Reconstructing roughly what remains from such archives could result to interesting studies of the long term evolution of these graphs. For large archives where one is interested in a particular type of pages, recrawling the full set of pages can be prohibitive. We propose to identify as quickly as possible a significant fraction of the still alive pages. Further selection can then be made to identify a set of pages suitable for the study and then to crawl them. Such techniques would be especially interesting when testing the liveness of an item is much lighter than downloading it completely. This is for instance the case for the Web with HEAD queries compared to GET queries. If a large amount of work has been devoted to maintaining fresh a set of crawled pages, little attention has been paid to the coverage of a partial recrawling a fairly old snapshot.

Second, some graphs tend to be harder to crawl with time. For instance, Twitter has continuously restricted its capacity to be crawled. Performing a full scan was possible a few years ago \cite{gabielkov:hal-00948889}, but it
can be prohibitively long nowadays. New techniques must thus be developed for identifying efficiently active accounts in such settings.

\subsection{Problem formulation}

Given an old snapshot, our goal is to identify a significant fraction of the items that are still alive or active now. The cost we incur is the number of fetches that are necessary to attain this a goal. A typical cost measure will be the average number of fetches per active item identified. The strategy for achieving this goal consists in producing an ordering for fetching the pages. We call \emph{LiveRank} an ordering such that the items that are still alive tend to appear first. We consider the problem of finding an efficient LiveRank in three settings: static when it is computed solely from the snapshot and the link relations recorded at that time, sampling-based when a sampling is performed in a first phase allowing to adjust the ordering according to the liveness of sampled items, or finally dynamic when it is incrementally computed as pages are fetched.

\subsection{Contribution}

We propose various LiveRank algorithms based on the graph structure of the snapshot. We evaluate them on two Web snapshots (from 10 to 20 millions nodes) and on a Twitter snapshot (40 million nodes). We propose several propositions based on the graph structure of the snapshot. We show that a rather simple combination of a small sampling phase and PageRank-like propagation in the remaining of the snapshot allows to gather from 15\% to 75\% of the active nodes with a cost that remains within a factor of 2 from the optimal ideal solution.

\subsection{Related work}

The process of crawling the Web has been extensively studied. A survey is given by Olston and Najork \cite{olston2010web}.

We focus here on batch crawling where the process starts from a given set of pages and terminates at some point.

This is classically opposed to incremental crawling where pages are continuously fetched. In incremental crawling, one of the main tuning is to balance between fetching new pages and refreshing old ones: the former increases coverage while the latter increases freshness. Both types may allow to discover new links towards unknown new pages (old pages can change). 
Cho and Garcia-Molina have extensively studied the problem of incremental crawling. For example, in \cite{cho2003effective}, they propose one of the first formalization of freshness and a thorough study on refreshing policies. They show the counter-intuitive result that adapting the frequency of crawl proportionally to the frequency of change works poorly with respect to the overall freshness of the fetched copy of the Web. The results have been extended with more elaborated variations of freshness. For instance, information longevity \cite{olston2008recrawl} considers the evolution of fragments of the content of a page.

The issue we investigate here is closer to a problem introduced by Cho and Ntoulas~\cite{cho2002effective}: they use sampling to estimate the frequency of change per site and then to fetch a set of pages such that the overall change ratio of the set is maximized. Their technique consists in estimating the frequency of page change per site and to crawl first sites with high frequency change. 
Tan et al. \cite{tan2007clustering} improve slightly over this technique by clusterizing the pages according to several features: not only their site (and other features read from the URL) but also content based features and linkage features (including PageRank and incoming degree). A change ratio per cluster is then estimating through sampling and clusters are downloaded in descending order of the estimated values.
More recently, Radinsky and Bennett~\cite{radinsky2013predicting} investigate a similar approach using learning techniques and avoiding the use of sampling.

Note that these approaches mainly focus on highly dynamic pages and use various information about pages whereas we are interested in stable nodes and we use only the graph structure, which is lighter.

With a slightly different objective, Dasgupta et al.~\cite{dasgupta2007discoverability} investigate how to discover new pages while minimizing the average number of fetches per new page found. Their work advocates for a greedy cover heuristic when a small fraction of the new pages has to be discovered quickly. On the opposite, they recommend a heuristic based on out-degrees for gathering a large fraction of the new pages. Their framework is close to ours and inspired the cost function used in this paper.

A related problem consist in estimating which pages are really valid among
the ``dangling'' pages on the frontier of the crawled web (those that are
pointed by crawled pages but that were not crawled themselves). Eiron et al.
propose to take this into account in the PageRank computation~\cite{eiron2004ranking}. In a similar trend, Bar-Yossef et al.~\cite{baryossef2004sic} propose
to compute a ``decay'' score for each page by refining on the proportion of
dead links in a page. Their goal is to identify poorly updated pages.
This score could be an interesting measure for computing a LiveRank, however
its computation requires the identification of dead links. It is thus not clear
how to both estimate it and at the same time try to avoid testing the liveness of possibly many dead pages.

While recrawling policies have been extensively studied for Web graph, other sources of online data such as social networks are less covered. Yet, it is possible to similarly crawl such networks. For example, one can explore the Twitter network by fetching information about user accounts that are linked by the follower-followee relations. However, crawling is much more restricted as all the data is possessed by a single company. 
This makes our approach even more relevant in such contexts where gathering a large amount can be extensively long.

Interestingly, Kwak et al.~\cite{kwak2010twitter} show, among various observations, a correlation between number of followers and PageRank. On the other hand the activity of a user measured in number of tweets seems to be more correlated to his number of followees that his number of followers.
First reported Twitter crawls include~\cite{java2007we,krishnamurthy2008few,kwak2010twitter}.
Recently, Gabiekov \emph{et al.}~\cite{gabielkov2012complete,gabielkov:hal-00948889} have presented a preliminary study on a complete picture of Twitter social graph.
The authors themselves claim that such extensive crawling becomes more and more difficult with time as Twitter tends to restrict its white list of IP authorized to query its API at high rate.

\subsection{Roadmap}

In the next Section, we propose a simple metric to evaluate the quality of a LiveRank and we introduce several classes of possible LiveRank solutions.
In Section \ref{sec:datasets}, we introduce three datasets, two from the \texttt{.uk} Web and one from Twitter, and we expose how a ground truth was computed for them.
Lastly, in Section \ref{sec:liveranks-evaluation}, we benchmark our LiveRanks against the datasets and discuss the results.

\section{Model}
\label{sec:model}

Let $ G=(V,E) $ be a graph obtained from a past crawl of a structured network. For example, $ G $ can represent:
\begin{itemize}
\item A Web graph, $ V $ representing the crawled pages and $ E $ the hyperlinks: for $ i $, $ j $ in $ V $, $ (i,j) $ is in $ E $ if, and only if, there is an hyperlink to $ j $ in $ i $. For Web graphs, edges are always directed;
\item A social network, $ V $ representing the users and $ E $ social relationships between them. For social networks, edges can be undirected (symmetric relationships like friendship) or directed (asymmetric relationship like follower/followee).
\end{itemize}

Let $ n $ denote the size of $ V $. At present time, only a subset of $ G $ is still active. The meaning of \emph{active} depends of the context and needs to be defined: alive pages for Web graphs, non-idle users for social networks, etc.
We call $ a $ the function that tells if nodes are active or not: $ a(X) $ denotes the active nodes from $ X\subset V $, while $ \bar{a}(X) $ stands for $ X \setminus a(X) $. Let $ n_a $ be $ |a(V)| $.

The problem we need to solve is can be expressed as: how to crawl a maximum number of pages from $ a(V) $ with a minimal crawling cost. In particular, one would like to avoid crawling too much pages from $ \bar{a}(V) $. If $ a $ was known, the task would be easy, but testing the activity of a node obviously requires to crawl it. This is the rationale for the notion of LiveRank.

\subsection{Performance metric}

Formally, any ordering can be seen as a LiveRank, so we need some performance metrics to define good LiveRanks that succeed in ranking the pages from $ a(V) $ first.
Following~\cite{dasgupta2007discoverability},
we define  the LiveRank cost as the average number of node retrievals necessary to obtain an active node, after a fraction $ 0< \alpha \leq 1 $ of the active nodes has been retrieved.

In details, let $ \mathcal{L}_i $ represent the $ i $ first pages returned by a LiveRank $ \mathcal{L} $, and let $ i(\mathcal{L},\alpha) $ be the smallest integer such that $ \frac{|a(\mathcal{L}_i)|}{n_a}\geq \alpha $. The cost function of $ \mathcal{L} $, which depends on $ \alpha $, is  then defined by:
$$\cost(\mathcal{L},\alpha)=\frac{i(\mathcal{L}, \alpha)}{\alpha n_a} \textbf{.}$$

A few remarks on the cost function:
\begin{itemize}
	\item
	It is always at greater than or equal to 1. An ideal LiveRank would perfectly separate $ a(V)$ from rest of the nodes, so its cost function would be $ 1 $. Without some oracle, this requires to test all pages, which is exactly what we would like to avoid. The cost function allows to capture this dilemma.
	\item
	Keeping a low cost becomes hard as $ \alpha $ gets close to $ 1 $: without some clairvoyant knowledge, capturing \emph{almost} all active nodes is almost as difficult as capturing all actives nodes. For that reason, one expects that when $ \alpha $ gets close to 1, the set of nodes any real LiveRank will need to crawl will tend to $ V $, leading to an asymptotical cost $ \frac{n}{n_a} $. This will be verified in Section \ref{sec:liveranks-evaluation}.
	\item
	Lastly, one may have noticed that the cost function uses $ n_a=|a(V)| $, for which an exact value requires a full knowledge of active nodes. This is not an issue here as we will perform our evaluation on datasets where $ a $ is known. For use on datasets without ground truth, one could either use an estimation of $ n_a $ based on a sampling or use a non-normalized cost function (for instance the fraction of active nodes obtained after $ i $ retrievals).
\end{itemize}

\subsection{PageRank}

Many of the LiveRanks proposed here are based on some variants of PageRank.
PageRank is a link analysis algorithm introduced in \cite{BP99} and used by the Google Internet search engine. It assigns a numerical importance to each page of a Web graph.
It uses the structural information from $ G $ to attribute importance according to the following (informal) recursive definition: \emph{a page is important if it is referenced by important pages}. Concretely, to compute PageRank value, denoted by the row vector $Y$, one needs to find the solution of the following equation:
\begin{eqnarray}
\label{eq:PR}
Y=dYA+(1-d)X\textbf{,}
\end{eqnarray}
\noindent where $A$ is a substochastic matrix derived from the adjacency matrix of $ G $, $d<1$ is a so-called damping factor (often set empirically to $d=0.85 $), and $ X \gvertneqq 0$ is a \emph{zap vector}. $ X $ represents a kind of importance \emph{by default} that is propagated from nodes to nodes according to $ A $ with a damping $ d $.

Computation of PageRank vectors has being widely studied. Several specific solutions were proposed and analysed \cite{LM04, BM05} including power method \cite{BP99}, adaptation \cite{ST03},  extrapolation \cite{TS03, CG03}, adaptive on-line method \cite{AP03}, etc.

We now present the different LiveRanks that we will consider in this paper. We broadly classify them in three classes: static, sample-based and dynamic.

\subsection{Static LiveRanks}

Static LiveRanks are computed offline using uniquely the information from $ G $. That makes them very basic, but also very easy to be used in a distributed way: 
given $ p $ crawlers of similar capacities, if $ \mathcal{L}=(l_1,\ldots,l_n) $, simply assign the task of testing node $l_i$ to crawler
$i\mod p$.

We propose the following three static LiveRanks. 

\paragraph{Random permutation ($ R $)} will serve both as a reference and as a building block for more advanced LiveRanks. $ R $ ignores any information from  $ G $, so its cost should be in average $\frac{n}{n_a}$, with a variance that tends to $ 0 $ as $ \alpha $ tends to $ 1 $. We expect good LiveRanks to have a cost function significantly lower than $ \cost(R) $.

\paragraph{Decreasing Indegree ordering ($ I $)} is a simple LiveRank that we expect to behave better than a random permutation. 
Intuitively, a high indegree can mean some importance, and important pages may be more robust. Also, older nodes should have more incoming edges (in terms of correlation), so high degree nodes can correspond to nodes that were already old at the time $ G $ was crawled, and old nodes may last longer than younger ones. Sorting by degree is the easiest way to exploit these correlations. 

\paragraph{PageRank ordering ($ P $)} pushes forward the \emph{indegree} idea. The intuition is that nodes that are still active are likely to point towards nodes that are still active also, even considering only old edges. This suggests to use a PageRank-like importance ranking. In absence of further knowledge, we propose to use the solution of \eqref{eq:PR} using $ d=.85 $ (typical value for Web graphs) and $X$ uniform on $ V $.

Note that it is very subjective to evaluate PageRank as an importance ranking, as importance should be ultimately validated by humans. On the other hand, the quality of PageRank as a static LiveRank is straightforward to verify, for instance using our cost metric.

The possible existence of correlation between Indegree (or PageRank) and activity will be investigated in Section \ref{sec:correlation}.

\subsection{Sample-based LiveRanks}

Using a LiveRank consists in crawling $ V $ in the prescribed order. During the crawl, the activity function $ a $ becomes partly available, and the obtained information could be used to enhance the retrieval.
Following that idea, we consider here a two-steps sample-based approach: we first fix a testing threshold $ z $ and test $ z $ items following a static LiveRank (like $ R $, $ I $ or $ P $). For the set $Z$ of nodes tested, called \emph{sample set} or \emph{training set}, $ a(Z) $ is known, which allows us to recompute the LiveRank of the remaining untested nodes.

Because the sampling uses a static LiveRank, and the adjusted new LiveRank is static as well, sample-based LiveRanks are still easy to use in a distributed way as the crawlers only need to receive crawl instructions on two occasions.

Notice that in the case where the sampling LiveRank is a random permutation, $ |a(Z)|\frac{n}{z} $ can be used as an estimate for $ n_a $. This can for instance be used to decide when to stop crawling if we desire to identify $ \alpha n_a $ active nodes in $ a(V) $.

\paragraph{Simple adaptive LiveRank $ (P_a) $}
When a node is active, we can assume it increases the chance that nodes it points to in $G$ are also active, and that activity is transmitted somehow through hyperlinks. Following this idea, a possible adaptive LiveRank consists in taking for $ X $ in \eqref{eq:PR} the uniform distribution on $ a(Z) $. This diffusion from such an initial set can be seen as a kind of breadth-first traversal starting from $a(Z)$, but with a PageRank flavour.

\paragraph{Double adaptive LiveRank $ (P_a^{+/-}) $}
The simple adaptive LiveRank does not use the information given by $\bar{a}(Z)$. One way to do this is to calculate an ``anti''-PageRank based on $\bar{a}(Z)$ instead of $a(Z)$. This ranking would represents a kind of diffusion of idleness, the underlying hypothesis being that idle nodes may point to nodes that tend to be idel too. As a result, we obtain a new LiveRank by combining these two PageRanks. After having tested several possible combinations not discussed in this paper, we empirically chose to weight each node by the ratio of the two sample-based PageRank, after having set all null entries of the anti-PageRank equal to the minimal non-null entry.

\paragraph{Active-site first LiveRank (ASF)}
To compare with previous work, we propose the following variant inspired by the Dasgupta et al.~\cite{dasgupta2007discoverability} strategy for finding pages that have changed in a recrawl. Their algorithm is based on sampling for estimating page change rate for each website and then to crawl sites by decreasing change rate.
In details, Active-site first (ASF) consists in partitioning $Z$ into websites determined by inspecting the URLs. We thus obtain a collection $Z_1,\ldots, Z_p$ of sets.
For each set $Z_i$ corresponding to some site $i$, we obtain an estimation
$|a(Z_i)| / |Z_i|$ of its activity (i.e. the fraction of active pages in the site).
We then sort the remaining URLs by decreasing site activity. Of course, this technique only works for Web graphs and can hardly be adjusted to other networks.

\subsection{Dynamic LiveRanks}

Instead of using the acquired information just one time after the sampling, Dynamic LiveRanks are continuously computed and updated on the fly along the entire crawling process. On the one hand, this gives them real-time knowledge of $ a $, but on the other hand, as the dynamic LiveRank may evolve all the time, they can create synchronization issues when used by distributed crawlers.

Like for sample-based LiveRanks, dynamic LiveRanks use a training set $ Z $ of $ z $ nodes from a static LiveRank. This allows to bootstrap the adjustment by giving a non-empty knowledge of $ a $, and prevents the LiveRank from focusing on only a small subset of $ V $.

\paragraph{Breadth-First Search (BFS)}
With BFS, we aim at taking direct advantage of the possible propagation of activity. The BFS queue is initialized with the (uncrawled) training set $ Z $. The next node to be crawled is popped from the queue following First-In-First-Out (FIFO) rule. If the selected node appears to be active, all of its uncrawled outgoing neighbors are pushed into the end of the queue. When the queue is empty, we pick the unvisited node with highest PageRank\footnote{We tested several other natural options and observed no significant impact.}.

\paragraph{Active indegree (AI)}
BFS uses a simple FIFO queuing to determine the processing order. We now propose AI which provides a more advanced node selection scheme. For AI, each node in the graph is associated with a \emph{activity score} value indicating how many reported active nodes point to it. These values are set to zeros at the beginning and always kept up-to-date. AI is initialized by testing $ Z $: each node in $a(Z)$ will increment the associated values of its out-going neighbors by one. After $ Z $ is tested, the next node to be crawled is simply the one with highest activity score (in case of equality, to keep things consistent, we pick the node with highest PageRank). Whenever a new active node is found, we update the activity scores of its untested neighbors.

\paragraph{}
With Dynamic LiveRank, it is natural to think of a last variant, a dynamic PageRank-based strategy where PageRank vector is recursively computed. Starting from a uniform distribution on $a(Z)$, we obtain $X$ in \eqref{eq:PR}. Then a new teleportation vector is constructed as a uniform distribution on largest value entries of $X$, \emph{i.e.,} those which are considered probably active after the first diffusion of $a(Z)$. The process continues and $X$ is updated iteratively. However, after some experimentations, we realized that this method is not efficient since it can not escape from the locality of $a(Z)$.

 \section{Datasets}
 \label{sec:datasets}
 
 We chose to evaluate the proposed LiveRanks on existing datasets of the Web and Twitter available on the WebGraph platform \cite{BoVWFI,BRSLLP,BCSU3}. In this Section, we present the datasets, describe how we obtained the activity function $ a $ and observe the correlations between $ a $, indegree and PageRank.
 
 \subsection{Webgraph Datasets}
 
For the study of Web graphs, we focused on snapshots of the British domain \texttt{.uk}.

\paragraph{\textbf{uk-2002} dataset}

The main dataset we use is the web graph \texttt{uk-2002}\footnote{\url{http://law.di.unimi.it/webdata/uk-2002/}} from UbiCrawler \cite{BCSU3}. This 2002 snapshot contains 18,520,486 pages and 298,113,762 hyperlinks.

The preliminary task is to determine $ a $, which is for Web graphs the liveness of the pages of the snapshot. 
For each URL, we have performed a GET request and hopefully obtained a corresponding HTTP code. Our main findings are:
\begin{itemize}
	\item One third of the total pages are no longer available today, the server returns error \texttt{404}.
	\item One fourth have a DNS problem (which probably means the website is also dead).
	\item For one fifth of the cases, the server sends back the redirection message \texttt{301}. Most redirections for pages of an old site lead to the root of a new site. If we look at the proportion of distinct pages alive at the end of redirections, it is as low as 0.1\%.
	\item Less than 13\% of pages return the code \texttt{200} (success). However, we found out that half of them actually display some text mentioning that the page was not found. To handle this issue, we have fully crawled all the pages of the dataset with code \texttt{200}
	and filtered out pages whose title or content have either \texttt{Page Not Found} or \texttt{Error 404}.
\end{itemize}

The results are summarized in Table~\ref{fig:stats_uk_2002}. In the end, our methodology led to finding out 1,164,998 alive pages, accounting for 6.4\% of the dataset.

\begin{table}[ht]
	\centering
	\begin{tabular}{|c|c|r|c|}
		\hline
		\textbf{Status}        &         \textbf{Description}         & \textbf{Number of pages} & \textbf{Percentage} \\ \hline\hline
		Code HTTP 404         &          Page not found             &                6 467 219 &       34,92\%        \\ \hline
		No answer          &           Host not found            &                4 470 845 &       24,14\%        \\ \hline
		Code HTTP 301         &             Redirection             &                3 455 923 &       18,66\%        \\
		\hdashline Target 301    & Target of redirection               &                   20 414 &        0,11\%        \\ \hline
		Code HTTP 200         &           Page exists             &                2 365 201 &       12,77\%        \\
		\hdashline True 200      & \textbf{Page really exists}       &                1 164 998 &   \textbf{6,29\%}    \\ \hline
		Others (403,\ldots)      &           Other error               &                1 761 298 &        9,51\%        \\ \hline
		Total             &           Graph size                 &               18 520 486 &        100\%         \\ \hline
	\end{tabular} 
	\caption{Status of web pages in \texttt{uk-2002}, crawled in December 2013.\label{fig:stats_uk_2002}}
\end{table}

\paragraph{\textbf{uk-2006} dataset}

The settings of \texttt{uk-2002} are rather adversarial (old snapshot with relatively few alive pages), so we wanted to evaluate the impact of LiveRanks on shorter time scales. In absence of fresh enough available datasets, we used the DELIS dataset~\cite{BSVLTAG}, 
a series of twelve continuous snapshots\footnote{\url{http://law.di.unimi.it/webdata/uk-union-2006-06-2007-05/}} starting from 06/2006 to 05/2007 (one-month intervals). We set $ G $ to the first snapshot (06/2006). It contains 31,316,403 nodes and 813,807,972 hyperlinks. We then considered the last snapshot (05/2007) as ``present time'',  setting the active set $a(V)$ as the intersection between the two snapshots. With this methodology, we hope to have a good approximation of $ a $ after a one-year period. For this dataset, we obtained $n_a=11,142,177$ ``alive'' nodes representing 35.56\% of the graph.

\subsection{Twitter Dataset}

Lastly, we used the dataset \texttt{twitter-2010} first introduced in \cite{kwak2010twitter}\footnote{\url{http://law.di.unimi.it/webdata/twitter-2010/}}. The graph contains roughly 42 millions Twitter user accounts and 1.5 billions follower-followee relationships among them. 
Arcs in the graph are directed from followers to followees: there is an arc from node $x$ to $y$ if user $x$ follows $y$. This orientation convention is in line with a PageRank approach: a user is important when she is followed by important users. Notice that information (tweets) traverses arcs in the opposite direction, from followees to followers.

We consider that a user is active if she has posted a tweet recently. For that purpose, we can query the Twitter interface to recover the timestamp of the last tweet of the user associated with a given identifier. 
Recovering the timestamps of all 41 millions users using Twitter API \cite{twitterAPI} would be extremely slow: when we made our measurements (05/2014), an authorized Twitter account was limited to 350 API requests/hour so querying all the accounts would have taken 13 years. While this is one of the main reasons for designing good LiveRank, we still need a full crawl to build a ground truth. To overcome this obstacle, we worked around the API limitation by using a browser-like crawler to recover each user timeline as if a regular browser was connecting to Twitter front servers. This is possible because the timestamp of the last entry can easily be scrapped from the HTML structure of the returned documents. However, such an approach becomes much more difficult for complex queries and might also be detected and prevented by Twitter in the future.

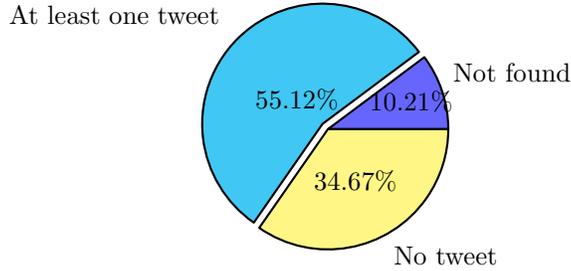
\begin{figure}[t]
	\centering
	\begin{tikzpicture}
	\pie[explode={0 , 0.1, 0}, radius=1.6] {10.21/ Not found , 55.12/At least one tweet, 34.67/No tweet}
	\end{tikzpicture}
	\caption{Statistics of the \texttt{twitter-2010} dataset}
	\label{fig:stats_Twitter2010}
\end{figure}

Having tested all nodes, we found three main categories of users corresponding to those who (i) no longer exist, (ii) have no tweet at all and (iii) have tweeted at least once before the crawling time. Figure~\ref{fig:stats_Twitter2010} shows the relative proportion of each category. 

For users with at least one tweet, we extracted the timestamp of their last tweet. After considering the cumulative distribution of last-tweet timestamps, we arbitrarily decided to set the activity threshold to six months: a user is active if she has tweeted during the last six months. With these settings, we obtained a list of 7,300,399 (17.53\%) active users, serving as ground truth for LiveRank evaluation.

\subsection{Correlations}\label{sec:correlation}

\begin{figure}[t]
	\centering
	\subfloat[Indegree CDF]{\label{fig:uk_CDF_Indegree}\includegraphics[width=0.47\textwidth]{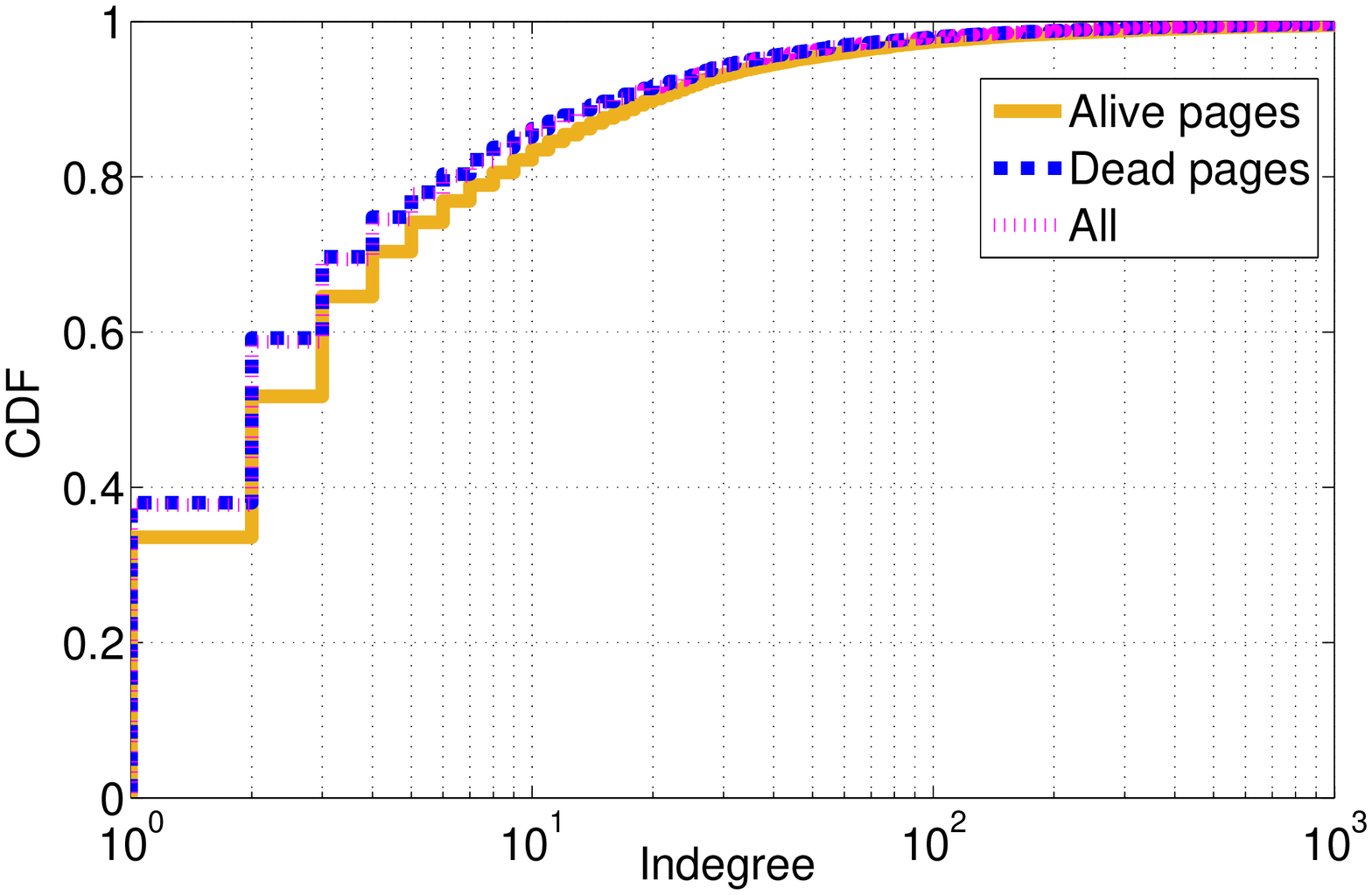}}
	\subfloat[PageRank CDF]{\label{fig:uk_CDF_PageRank}\includegraphics[width=0.47\textwidth]{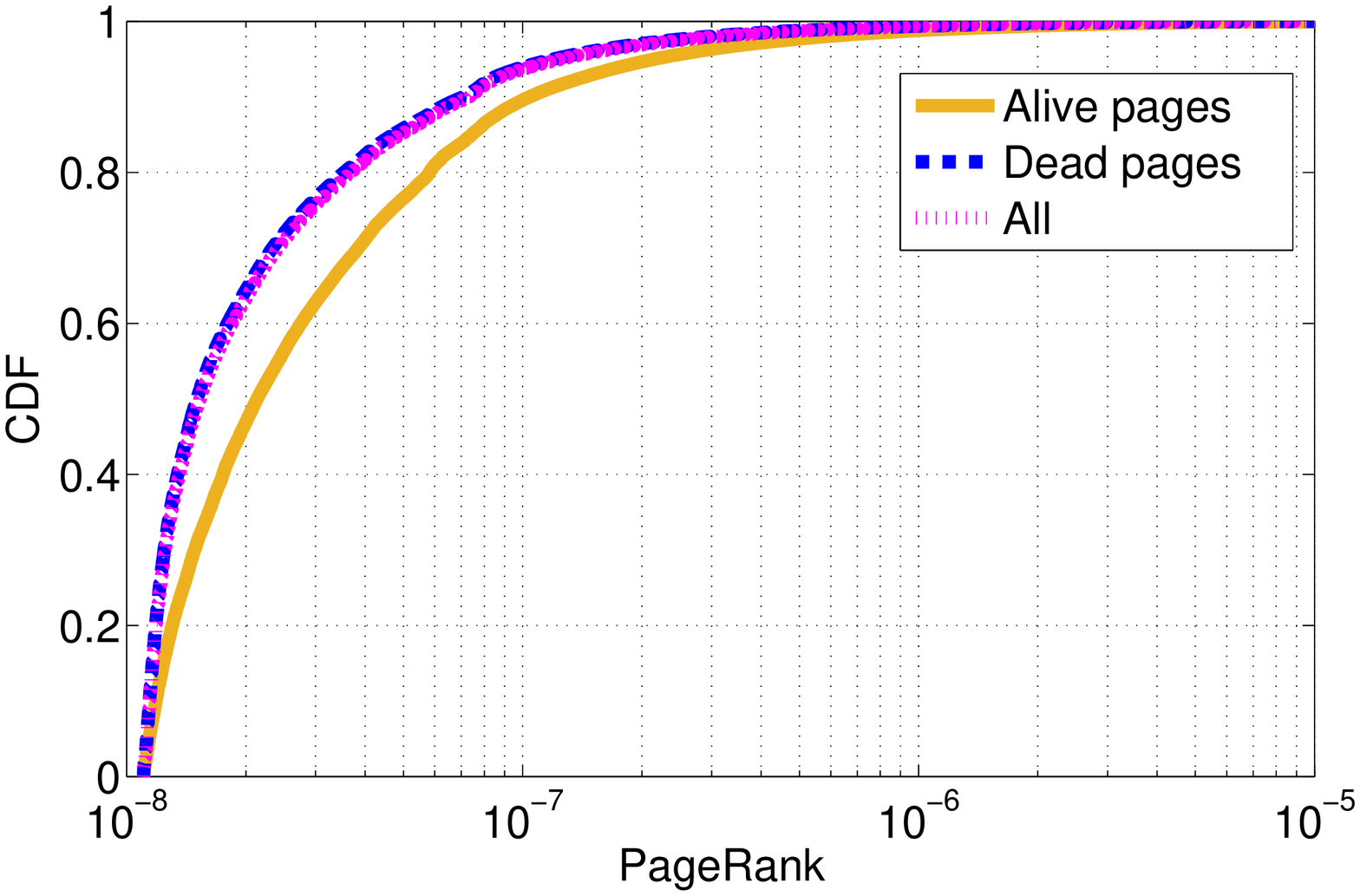}}
	\caption{Impact of liveness to Indegree/PageRank distribution for \texttt{uk-2002}.}
	\label{fig:CDFuk02}
\end{figure}
	
\begin{figure}[t]	
	\subfloat[Followers CDF]{\label{fig:twitter_CDF_Indegree}\includegraphics[width=0.47\textwidth]{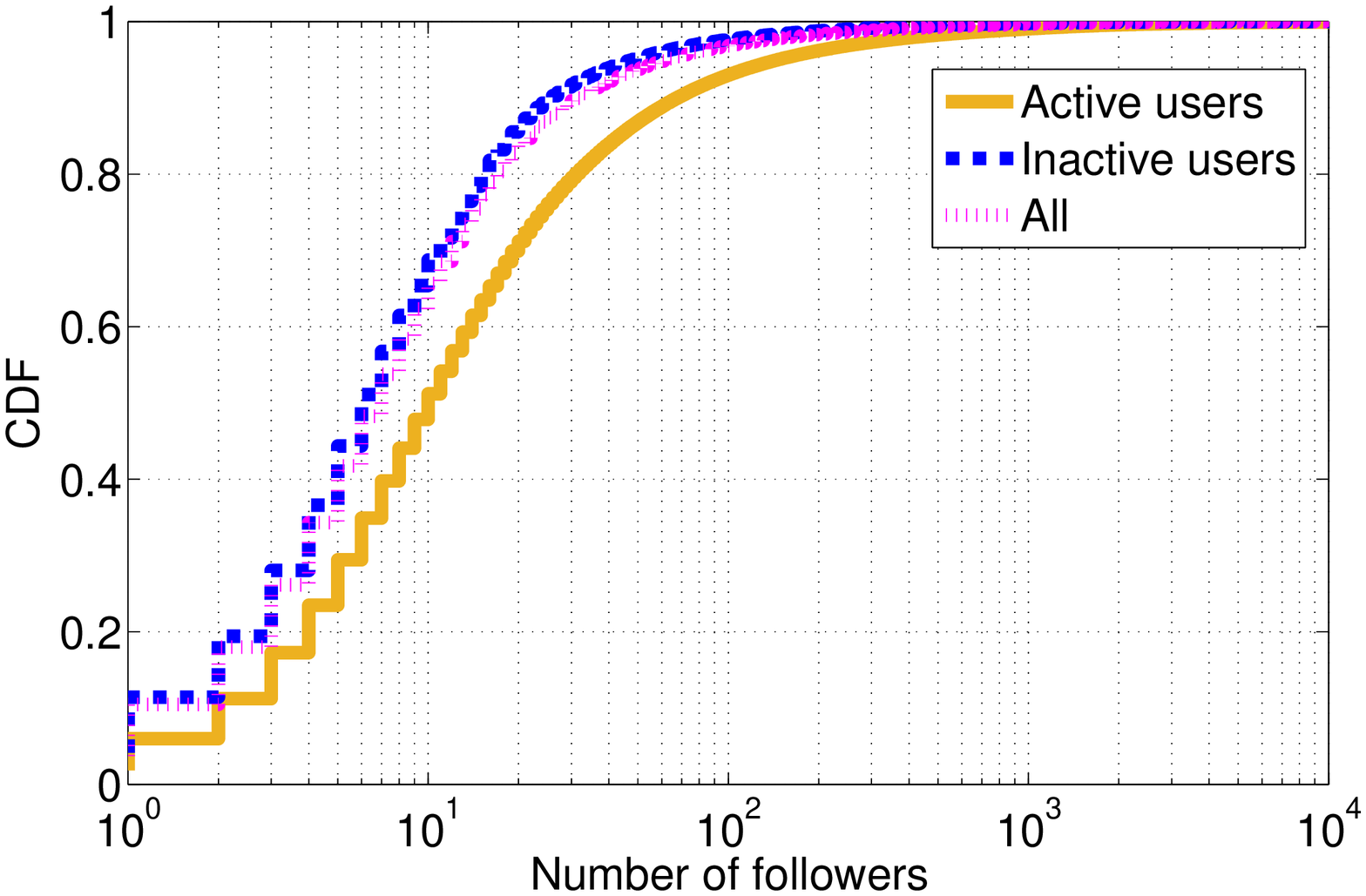}}
	\subfloat[PageRank CDF]{\label{fig:twitter_CDF_PageRank}\includegraphics[width=0.47\textwidth]{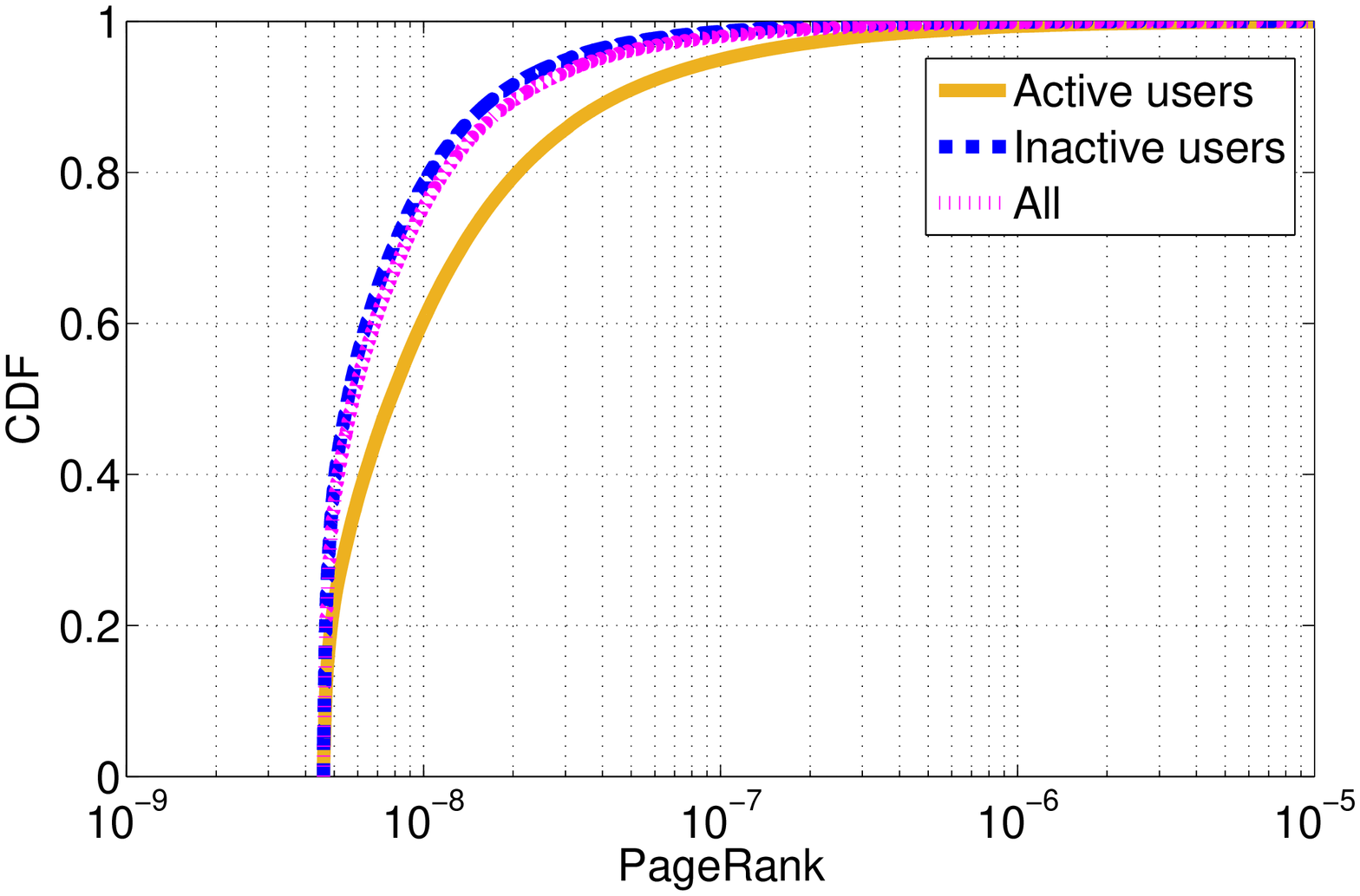}}
	\caption{Impact of activity to Followers/PageRank for \texttt{twitter-2010}.}
	\label{fig:CDFt10}
\end{figure}

The rationale behind the LiveRanks $ I$ and $ P $ is the assumption that the activity of nodes is correlated to the graph structure of the snapshot, so that a node with high in-degree or PageRank has more chances to stay active.

To validate this, we plot in Figure~\ref{fig:CDFuk02} the cumulative distribution of in-degree (figure \ref{fig:uk_CDF_Indegree}) and PageRank  (figure \ref{fig:uk_CDF_PageRank}) for alive, dead, and all pages of the \texttt{uk-2002} dataset. We observe that the curve for active nodes is slightly shifted to the right compared to the other curves in each figures: active users tend to have slightly higher in-degree and PageRank than in the overall population. The bias is bigger for PageRank, suggesting that LiveRank (P) should perform better than LiveRank (I) for Web graphs.

Figure \ref{fig:CDFt10} presents the same results for the \texttt{twitter-2010} dataset. While the curves are qualitatively similar, the bias comparison is not as clearly in favor of PageRank.

We will now measure how this bias impacts the cost function of corresponding LiveRanks.

\section{LiveRanks evaluation}
\label{sec:liveranks-evaluation}

After having proposed several LiveRanks  in Section \ref{sec:model} and described our datasets in previous Section, we can now benchmark our proposals.

All our evaluations are based on representations of the cost functions. In each plot, the x-axis indicates the fraction $\alpha$ of active nodes we aim to discover and the y-axis corresponds to the relative cost of the crawl required to achieve that goal. A low curve indicates an efficient LiveRank. Like said in Section \ref{subsec:cost}: an ideal LiveRank would achieve a constant cost of 1; a random LiveRank is quickly constant with an average cost $n/n_a$; any non-clairvoyant LiveRank will tend to cost $n/n_a$ as $ \alpha $ goes to 1.

\subsection{Evaluation on Web Graphs}

We mainly focus here on the \texttt{uk-2002} dataset. Unless otherwise specified, the training set contains the $ z=100 000 $ pages of higher (static) PageRank.

\subsubsection{Static and sample-based LiveRanks}

\begin{figure}[ht]
	\centering
	\includegraphics[width=0.65\textwidth]{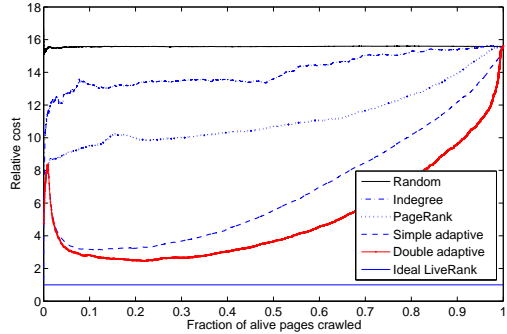}
	\caption{Main results on static and sample-based LiveRanks (\texttt{uk-2002})}
	\label{fig:uk_2002_Efficiency}
\end{figure}

We first evaluate the results of static and sample-based LiveRanks. The results are displayed in Figure~\ref{fig:uk_2002_Efficiency}.
For static LiveRanks, we see as expected that a random ordering gives an almost constant cost equal to $ \frac{n}{n_a}\approx 15.6 $. Indegree ordering (I) and PageRank (P) significantly outperform this result, PageRank being the best of the three: it is twice more efficient than random for small $\alpha$, and still performs approximately 30\% better when up to $\alpha=0.6$. 
We then notice that we can get even much better costs with sample-based approaches, the double-adaptive LiveRank $ P_a^{+/-} $ giving a significant improvement over the simple-adaptive one $ P_a $. $ P_a^{+/-} $ allows improving the ordering by a factor of 6 approximately around $\alpha=0.2$ with a cost of 2.5 fetches per active node found. The cost for gathering half of the alive pages is less than 4, and for $90\%$ it stays under 10.

\subsubsection{Quantitative and qualitative impact of the training set}

\begin{figure}[ht]
	\centering
	\subfloat[Impact of the size $z$]{\label{fig:uk_2002_z}\includegraphics[width=0.45\textwidth]{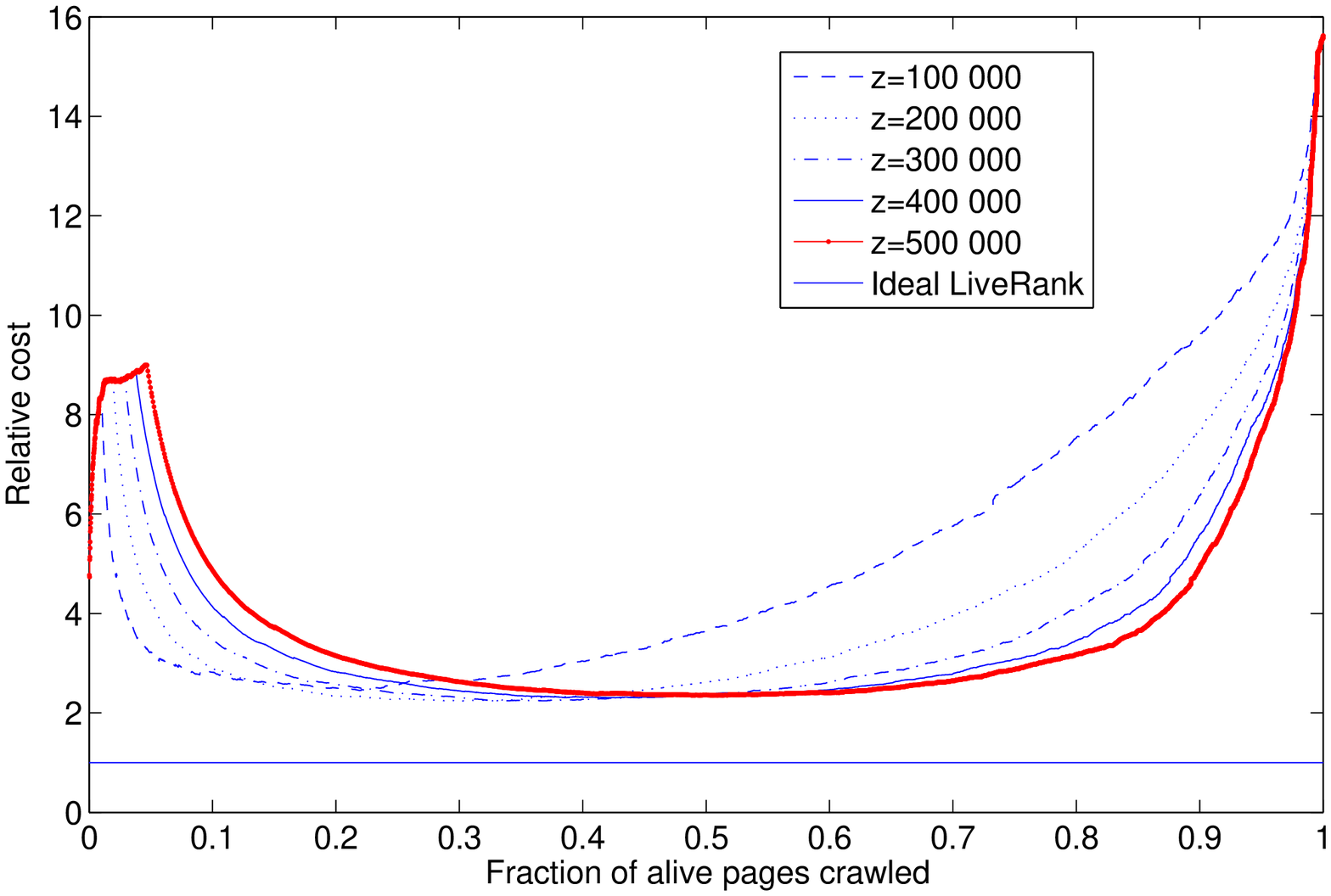}}
	\subfloat[Impact of the selection of $ Z $]{\label{fig:uk_2002_PRAlgo_ChangingSeeds}\includegraphics[width=0.45\textwidth]{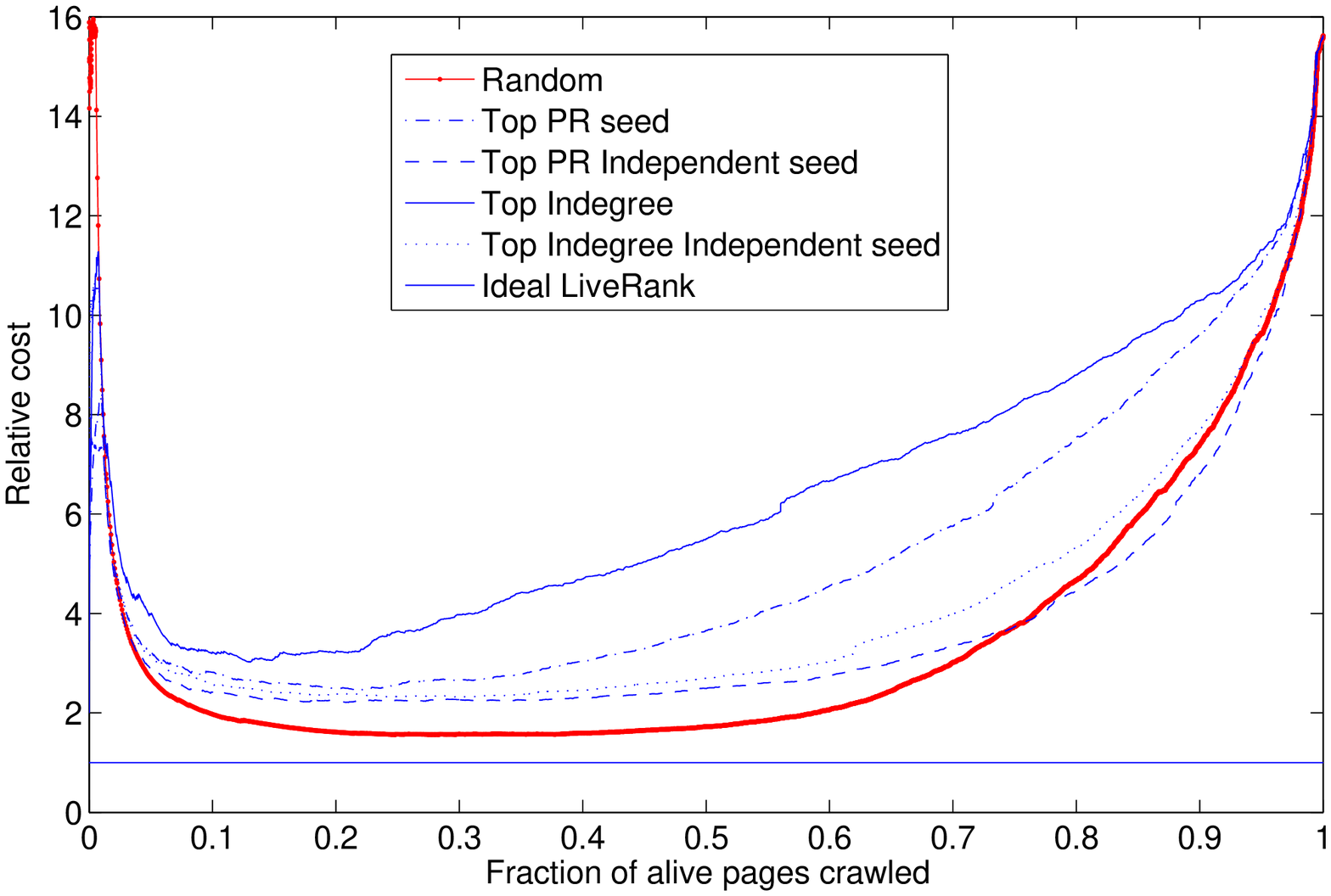}}
	\caption{Impact of the training set (\texttt{uk-2002})}
	\label{fig:training}
\end{figure}

We study in Figure \ref{fig:training} the impact of the training sets on sample-based LiveRanks. Results are shown for $ P_a^{+/-} $ but similar results were obtained for $ P_a $.

Figure~\ref{fig:uk_2002_z} shows the impact of the size $z$ of the sampling set (sampling the top PageRank pages). We observe some trade-off: as the sampling set grows larger, the initial cost increases as the sample does not use any fresh information, but it results in a significant increment of efficiency in the long run. For this dataset, taking a big training set ($z$=500 000) allows reducing the cost of the crawl for $\alpha \geq 0.4$, and maintains a cost less than 4 for up to 90\%. 

Another key aspect of the sampling phase is the qualitative choice of the sample set. Using $z$=100 000, we can observe in Figure~\ref{fig:uk_2002_PRAlgo_ChangingSeeds} that the performance of double adaptive $ P_a^{+/-} $ is further improved by using a random sample set rather than selecting it according to the PageRank or by decreasing indegree. We believe that the reason is that a random sample avoids a locality effect in the sampling set as high PageRank pages tend to concentrate in some local parts of the graph. To verify that, we tried to modify Indegree and PageRank selection to avoid the selection of neighbor pages. The results (not displayed here) show a significant improvement while staying less efficient than using a random sample.

To summarize, double-adaptive LiveRank through random sampling seems to offer a very low cost, within a factor of 2 from optimal for a large range of values $\alpha$.

\subsubsection{Dynamic LiveRanks}

\begin{figure}[ht]
	\centering
	\subfloat[Comparison with double adaptive]{\label{fig:uk_2002_PR-BFS-Indeg}\includegraphics[width=0.45\textwidth]{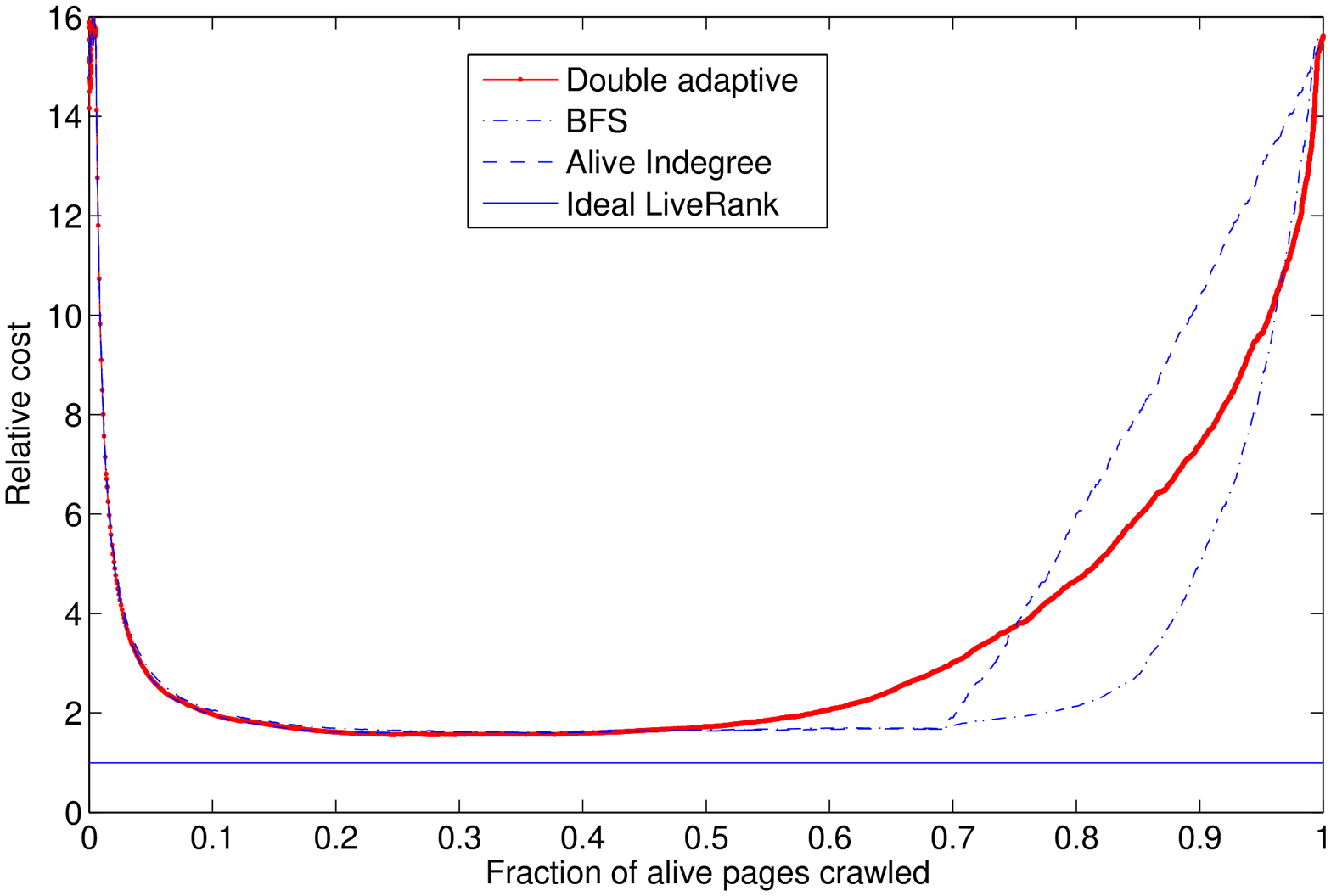}}
	\subfloat[Impact of $ Z $ on dynamic LiveRanks]{\label{fig:uk_2002_BFS-Indeg_ChangingSeeds}\includegraphics[width=0.45\textwidth]{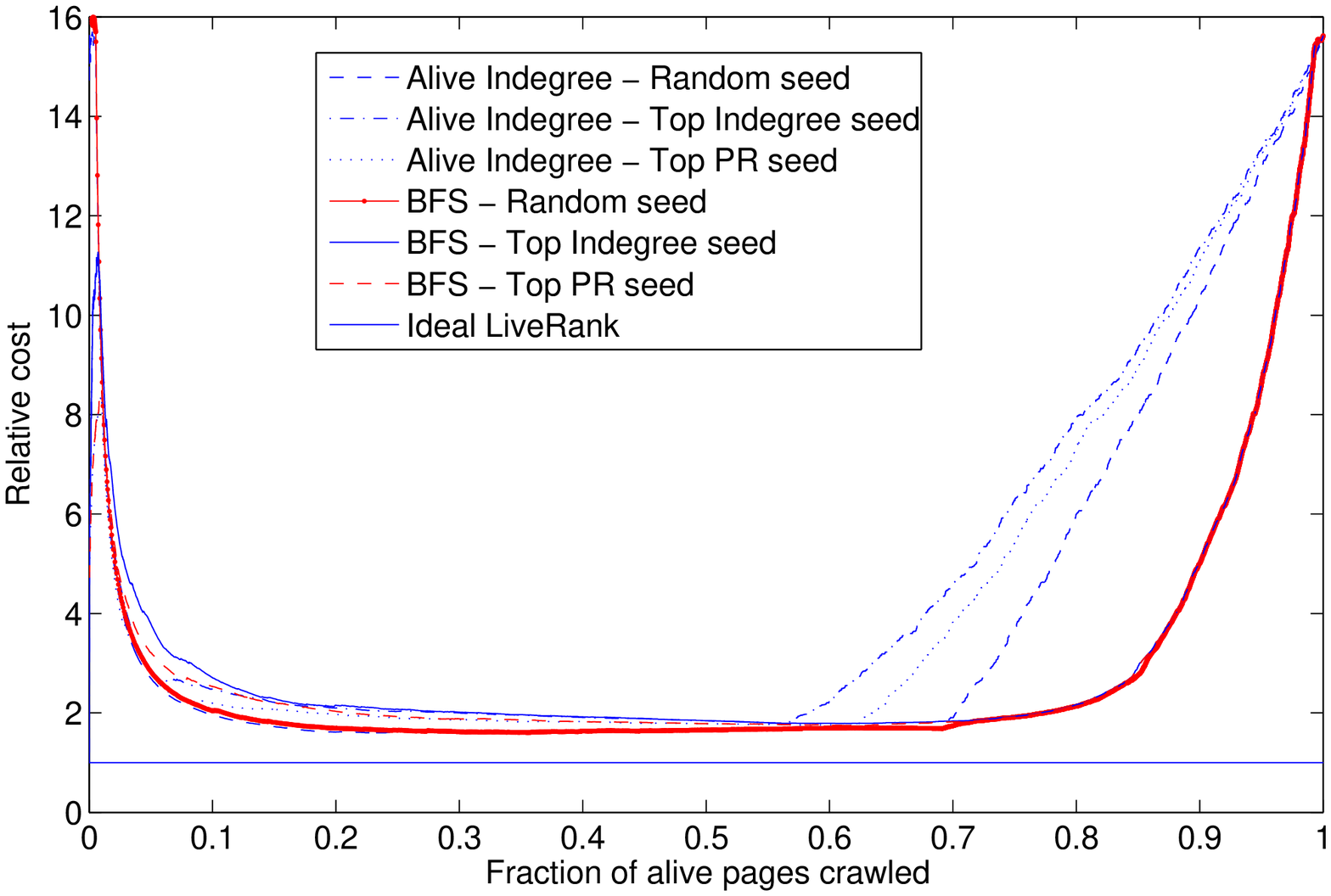}}
	\caption{Performance of dynamic LiveRanks (\texttt{uk-2002})}
	\label{fig:uk2002}
\end{figure}

We then consider the performance of fully dynamic strategies, using the double-adaptive LiveRank with random training set as a landmark. The results are displayed in Figure~\ref{fig:uk_2002_PR-BFS-Indeg}. We see that bread-first search BFS and active indegree AI perform similarly to double adaptive $ P_a^{+/-} $ for low $\alpha$ and can outperform it for large $\alpha$ (especially BFS).
BFS begin to significantly outperform double adaptive for $\alpha\geq 0.5$. However, if one needs to gather half of the active pages or less, double adaptive is still the best candidate as it is much simpler to operate, especially with a distributed crawler.

Additionally, Figure~\ref{fig:uk_2002_BFS-Indeg_ChangingSeeds} shows the impact of different sampling sets on BFS and AI. Except for high values of $ \alpha $ where a random sampling outperforms other strategies, the type of sampling does not seem to affect the two dynamic LiveRanks as much as it was observed for the double-adaptive LiveRank.

\subsubsection{\texttt{uk-2006} dataset}

\begin{figure}[ht]
	\centering
	\includegraphics[width=0.65\textwidth]{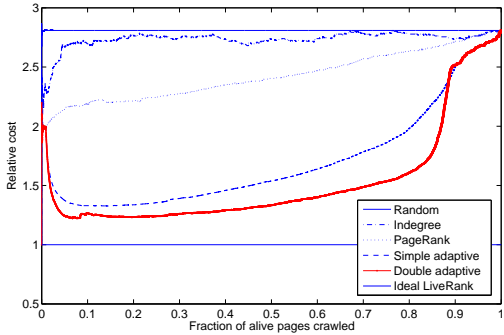}
	\caption{\texttt{uk-2006} main evaluation results}
	\label{fig:uk2006}
\end{figure}

We have repeated the same experiments on the dataset \texttt{uk-2006}, where the update interval is only one year. Figure~\ref{fig:uk2006} shows the results for static and sample-based LiveRanks, using $z$=200 000 (because the dataset is larger) and random sampling. The observation are qualitatively quite similar to \texttt{uk-2002}. The main difference is that all costs are lower due to a higher proportion of alive pages ($ \frac{n}{n_a}\approx 2.81 $). The double-adaptive version still gives the lower relative cost among static and sample-based LiveRanks, staying under 1.4 for a wide range of $ \alpha $.

\subsubsection{Comparison with a site-based approach}

\begin{figure}[t]
	\centering
	\includegraphics[width=.65\textwidth]{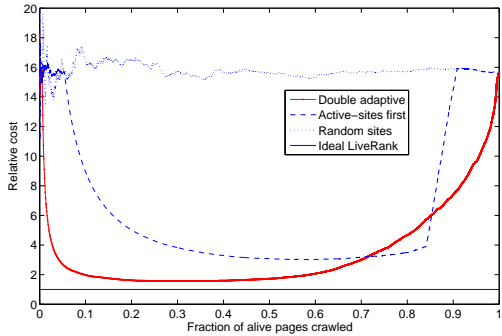}
	\caption{Comparison with active-site first LiveRank (\texttt{uk-2002})}
	\label{fig:SammplingAlgo}
\end{figure}

To benchmark with techniques from previous work for finding web pages that been updated after a crawl, Figure~\ref{fig:SammplingAlgo} compares double adaptive $ P_a^{+/-} $ to active-site first ASF with random sampling. The number of random pages tested in each site and the overall number of tests are the same for both methods. Note that given the budget $z$, it was not possible to sample small websites. Unsampled websites are crawled after the sampled ones.

We see that for $ \alpha $ greater than $ 0.9 $, ASF performs like a random LiveRank. This corresponds to the point where all sampled website have been crawled. That effect aside, the performance of ASF is not as good as double-adaptive LiveRank for earlier $ \alpha $. In the end, ASF only beats $ P_a^{+/-} $ for a small range of $ \alpha $, between 0.7 and 0.85, and the gain within that range stays limited.

\subsection{Evaluation on Twitter}

As discussed before, the Twitter graph has structural properties distinct from Web graphs. In this part we analyze how these differences affect the performance of LiveRanks.

\subsubsection{Static and sampled-based LiveRanks}

\begin{figure}[ht]
	\centering
	\includegraphics[width=0.65\textwidth]{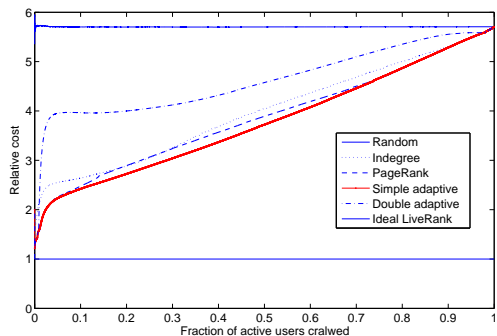}
	\caption{Main results on static and sample-based LiveRanks (\texttt{twitter-2010})}
	\label{fig:twitter_2010_Efficiency}
\end{figure}

Figure~\ref{fig:twitter_2010_Efficiency} compares the static and sample-based LiveRanks. A first observation is that the double adaptive LiveRank $ P_a^{+/-} $ performs very poorly compared to the other LiveRanks, including Indegree \emph{I}. It indicates that if the intuition of some death propagation was relevant for Web graphs (it was a convenient way to spot dead web sites for instance), this is not the case for Twitter: the fact that followers become inactive does not seem to have an impact on the activity of the followees.
In the end, the simple adaptive LiveRank $ P_a $ has the best performance, closely followed by the static LiveRanks $ P $ and $ I $. The three of them have a cost function that seem to grow roughly linearly between 2 and 4 as $ \alpha $ goes from 0 to 0.6.

\subsubsection{Quantitative and qualitative impact of the training set}

\begin{figure}[ht]
	\centering
	\subfloat[Impact of the size $z$]{\label{fig:twitter_2010_z}\includegraphics[width=0.45\textwidth]{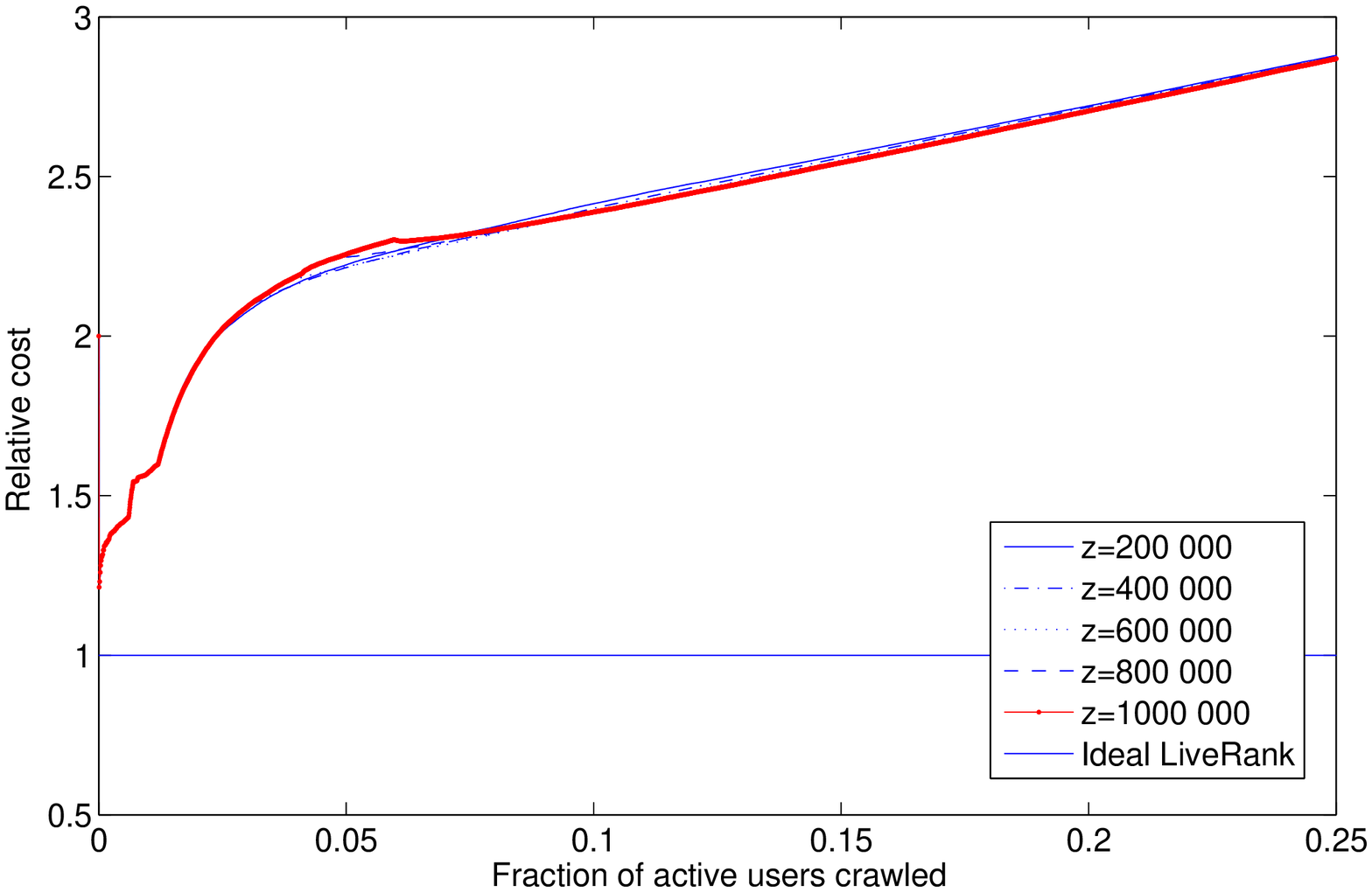}}
	\subfloat[Impact of the selection of $ Z $]{\label{fig:twitter_2010_PRAlgo_ChangingSeeds}\includegraphics[width=0.45\textwidth]{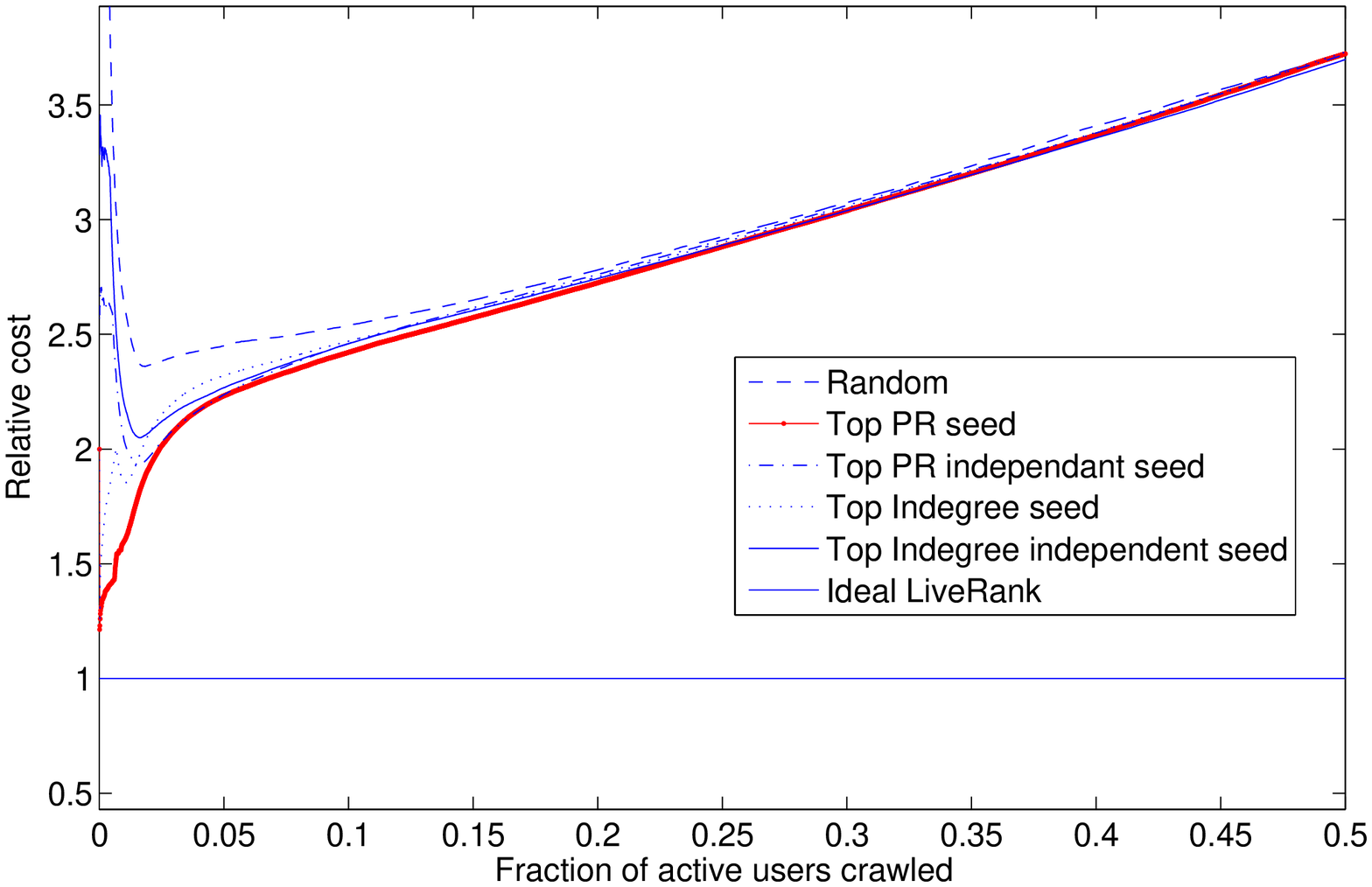}}
	\caption{Impact of the training set (\texttt{twitter-2010})}
	\label{fig:trainingrwitter}
\end{figure}

In Figure~\ref{fig:twitter_2010_z}, we vary the size of the training set, ranging from $z=$200 000 to $z=$1000 000.  Results indicate that the cost function is almost not affected by $ z $  as long as it is high enough. Compared to the results observed on Web graphs, this means that taking a big training set: (i) will not burden the cost function for small $ \alpha $. This likely comes from the fact that the sampling set is PageRank-based by default, and the static PageRank is already close to the best LiveRank we can get; (ii) will not improve the performance for large $ \alpha $ either, meaning that no significantly useful knowledge is obtained after some threshold. This relative independence with respect to $ z $ is another qualitative difference compared to Web graphs.

Figure~\ref{fig:twitter_2010_PRAlgo_ChangingSeeds} shows the impact of training set types on simple adaptive LiveRank $P_a$. Unlike Web graphs where random sampling dominates others, in social network the training set filled by PageRank is the best whereas the random seed is worse. This can be interpreted as a result of a weaker structural locality (\emph{i.e.,} no highly correlated clusters like web sites for Web graphs), so that activeness is more concentrated around important Twitter individual users that should be considered as soon as possible.

\subsubsection{Dynamic LiveRanks}

\begin{figure}[ht]
	\centering
	\subfloat[Comparison with double adaptive]{\label{fig:twitter_2010_PR-BFS-Indeg}\includegraphics[width=0.45\textwidth]{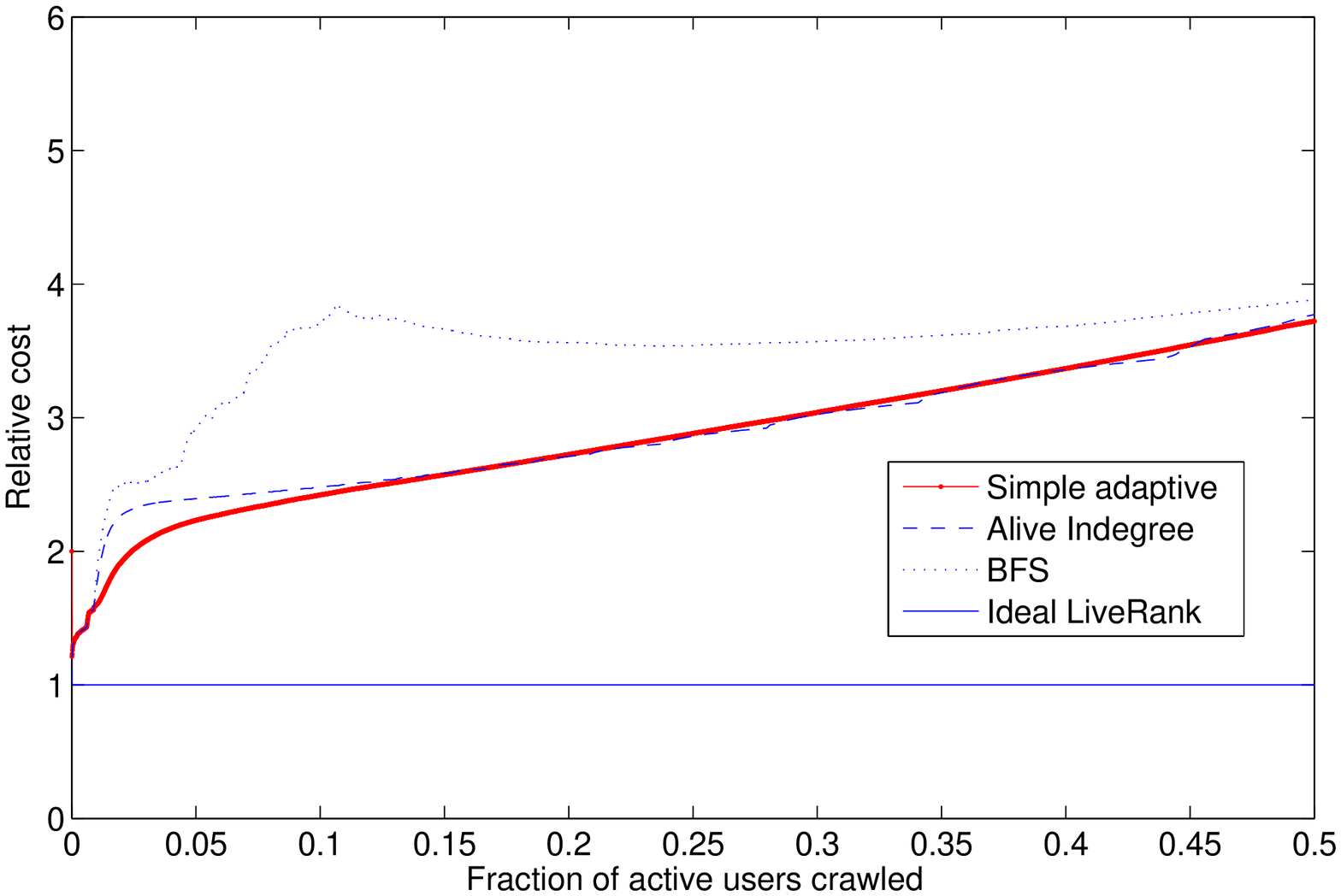}}
	\subfloat[Impact of $ Z $ on dynamic LiveRanks]{\label{fig:twitter_2010_BFS-Indeg_ChangingSeeds}\includegraphics[width=0.45\textwidth]{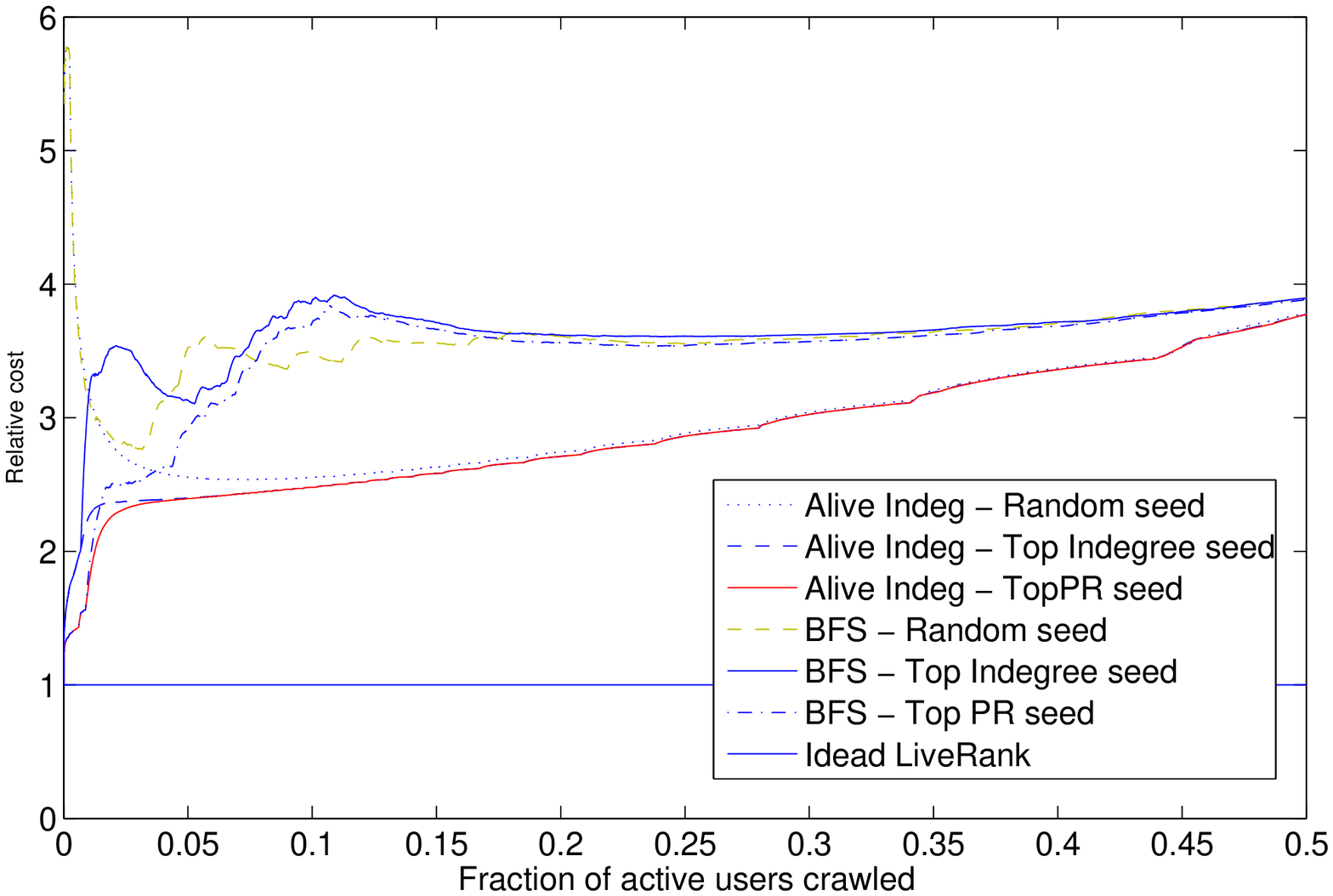}}
	\caption{Performance of dynamic LiveRanks (\texttt{twitter-2010})}
	\label{fig:dyntwit}
\end{figure}

In Figure~\ref{fig:twitter_2010_PR-BFS-Indeg}, we compare the simple adaptive PageRank $ P_a $ with the dynamic LiveRanks. All of them are initialized with default values (PageRank sampling of size $z=$100 000). $ P_a $ stays the best option: it is slightly better than AI and much more efficient than BFS. While for Web graphs, dynamics LiveRanks could still be preferred for some settings, it seems that in the context of Twitter it is never the case especially considering their deploiement complexity in a distributed crawler.

Lastly, Figure~\ref{fig:twitter_2010_BFS-Indeg_ChangingSeeds} indicates the impact of different training sets on the two dynamic LiveRanks. It confirms that the combination of AI and a PageRank-ordered training set gives the best results for that type of LiveRanks, which is still not enough to compete against $ P_a $.

\section{Conclusion}

In this paper, we investigated how to efficiently retrieve large portions of alive pages from an old crawl using orderings we called LiveRanks.
We observed that PageRank is a good static LiveRank, 
which can be significantly improved by first testing a small fraction of the pages for adjustment in a sample-based approach.

Compared to previous work on identifying modified pages, our technique performs similarly for a given large desired fraction (around 80\%) when compared to the LiveRank algorithm inspired by the technique in \cite{cho2002effective}. However, outside that range, our method outperforms this technique. Interesting future work could reside in using our techniques for the problem exposed in \cite{cho2002effective} (identification of pages that have changed) and compare with the Website sampling approach.

Another advantage of our technique is the possibility to be applied to more general types of structured networks, like Twitter. However, it seems that the choice of an appropriate LiveRank is closely related to the type of network.

Interestingly, we could not get significant gain when using fully dynamic LiveRanks. As noticed before, each of the two phases of the sample-based approach can be easily parallelized through multiple crawlers whereas this would be much more difficult with a fully dynamic approach. The sample-based method could for example be implemented with in two rounds of a simple map-reduce program whereas the dynamic approach requires continuous exchanges of messages between the crawlers.

Our work establishes the possibility of efficiently recovering a significant portion of the active nodes of an old snapshot and advocates for the use of an adaptive sample-based PageRank for obtaining an efficient LiveRank.

To conclude, we emphasize again that the LiveRank approach proposed in this paper is very generic, and its field of applications is not limited to Web graphs or Twitter. It can be straightforwardly adapted to any online data with similar linkage enabling crawling.

\paragraph{Acknowledgments}
The work presented in this paper has been carried out at LINCS (\url{http://www.lincs.fr}) and has been founded by the European project Reveal (\url{http://revealproject.eu/}).

\bibliographystyle{abbrv}

\begin{thebibliography}{}

\end{thebibliography}


\begin{thebibliography}{10}

\bibitem{twitterAPI}
Twitter graph 2010.
\newblock \url{https://dev.twitter.com/}.

\bibitem{AP03}
S.~Abiteboul, M.~Preda, and G.~Cobena.
\newblock Adaptive on-line page importance computation.
\newblock In {\em Proceedings of the 12th International Conference on World
  Wide Web}, WWW '03, pages 280--290, New York, NY, USA, 2003. ACM.

\bibitem{baryossef2004sic}
Z.~Bar-Yossef, A.~Z. Broder, R.~Kumar, and A.~Tomkins.
\newblock Sic transit gloria telae: Towards an understanding of the web's
  decay.
\newblock In {\em WWW '04}, pages 328--337, 2004.

\bibitem{BM05}
M.~Bianchini, M.~Gori, and F.~Scarselli.
\newblock Inside pagerank.
\newblock {\em ACM Trans. Internet Technol.}, 5(1):92--128, Feb. 2005.

\bibitem{BCSU3}
P.~Boldi, B.~Codenotti, M.~Santini, and S.~Vigna.
\newblock Ubicrawler: A scalable fully distributed web crawler.
\newblock {\em Software: Practice \& Experience}, 34(8):711--726, 2004.

\bibitem{BRSLLP}
P.~Boldi, M.~Rosa, M.~Santini, and S.~Vigna.
\newblock Layered label propagation: A multiresolution coordinate-free ordering
  for compressing social networks.
\newblock In {\em Proceedings of the 20th international conference on World
  Wide Web}. ACM Press, 2011.

\bibitem{BSVLTAG}
P.~Boldi, M.~Santini, and S.~Vigna.
\newblock A large time-aware graph.
\newblock {\em SIGIR Forum}, 42(2):33--38, 2008.

\bibitem{BoVWFI}
P.~Boldi and S.~Vigna.
\newblock The {W}eb{G}raph framework {I}: {C}ompression techniques.
\newblock In {\em Proc. of the Thirteenth International World Wide Web
  Conference (WWW 2004)}, pages 595--601, Manhattan, USA, 2004. ACM Press.

\bibitem{cho2003effective}
J.~Cho and H.~Garcia-Molina.
\newblock Effective page refresh policies for web crawlers.
\newblock {\em ACM Transactions on Database Systems (TODS)}, 28(4):390--426,
  2003.

\bibitem{cho2002effective}
J.~Cho and A.~Ntoulas.
\newblock Effective change detection using sampling.
\newblock In {\em Proceedings of the 28th international conference on Very
  Large Data Bases}, pages 514--525. VLDB Endowment, 2002.

\bibitem{dasgupta2007discoverability}
A.~Dasgupta, A.~Ghosh, R.~Kumar, C.~Olston, S.~Pandey, and A.~Tomkins.
\newblock The discoverability of the web.
\newblock In {\em Proceedings of the 16th international conference on World
  Wide Web}, pages 421--430. ACM, 2007.

\bibitem{eiron2004ranking}
N.~Eiron, K.~S. McCurley, and J.~A. Tomlin.
\newblock Ranking the web frontier.
\newblock In {\em Proceedings of the 13th international conference on World
  Wide Web}, pages 309--318. ACM, 2004.

\bibitem{gabielkov2012complete}
M.~Gabielkov and A.~Legout.
\newblock The complete picture of the twitter social graph.
\newblock In {\em Proceedings of the 2012 ACM conference on CoNEXT student
  workshop}, pages 19--20. ACM, 2012.

\bibitem{gabielkov:hal-00948889}
M.~Gabielkov, A.~Rao, and A.~Legout.
\newblock {Studying Social Networks at Scale: Macroscopic Anatomy of the
  Twitter Social Graph}.
\newblock In {\em {ACM Sigmetrics 2014}}, Austin, United States, June 2014.

\bibitem{TS03}
T.~Haveliwala, A.~Kamvar, D.~Klein, C.~Manning, and G.~Golub.
\newblock Computing pagerank using power extrapolation.
\newblock Technical report, 2003.

\bibitem{java2007we}
A.~Java, X.~Song, T.~Finin, and B.~Tseng.
\newblock Why we twitter: understanding microblogging usage and communities.
\newblock In {\em Proceedings of the 9th WebKDD and 1st SNA-KDD 2007 workshop
  on Web mining and social network analysis}, pages 56--65. ACM, 2007.

\bibitem{ST03}
S.~Kamvar, T.~Haveliwala, and G.~Golub.
\newblock Adaptive methods for the computation of pagerank.
\newblock Technical Report 2003-26, Stanford InfoLab, April 2003.

\bibitem{CG03}
S.~D. Kamvar, T.~H. Haveliwala, C.~D. Manning, and G.~H. Golub.
\newblock Extrapolation methods for accelerating pagerank computations.
\newblock In {\em Proceedings of the 12th International Conference on World
  Wide Web}, WWW '03, pages 261--270, New York, NY, USA, 2003. ACM.

\bibitem{krishnamurthy2008few}
B.~Krishnamurthy, P.~Gill, and M.~Arlitt.
\newblock A few chirps about twitter.
\newblock In {\em Proceedings of the first workshop on Online social networks},
  pages 19--24. ACM, 2008.

\bibitem{kwak2010twitter}
H.~Kwak, C.~Lee, H.~Park, and S.~Moon.
\newblock What is twitter, a social network or a news media?
\newblock In {\em Proceedings of the 19th international conference on World
  wide web}, pages 591--600. ACM, 2010.

\bibitem{LM04}
A.~N. Langville and C.~D. Meyer.
\newblock Deeper inside pagerank.
\newblock {\em Internet Mathematics}, 1:2004, 2004.

\bibitem{olston2010web}
C.~Olston and M.~Najork.
\newblock Web crawling.
\newblock {\em Foundations and Trends in Information Retrieval}, 4(3):175--246,
  2010.

\bibitem{olston2008recrawl}
C.~Olston and S.~Pandey.
\newblock Recrawl scheduling based on information longevity.
\newblock In {\em Proceedings of the 17th international conference on World
  Wide Web}, pages 437--446. ACM, 2008.

\bibitem{BP99}
L.~Page, S.~Brin, R.~Motwani, and T.~Winograd.
\newblock In {\em The PageRank Citation Ranking: Bringing Order to the Web.},
  number 1999-66. Stanford InfoLab, November 1999.

\bibitem{radinsky2013predicting}
K.~Radinsky and P.~N. Bennett.
\newblock Predicting content change on the web.
\newblock In {\em Proceedings of the sixth ACM international conference on Web
  search and data mining}, pages 415--424. ACM, 2013.

\bibitem{tan2007clustering}
Q.~Tan, Z.~Zhuang, P.~Mitra, and C.~L. Giles.
\newblock A clustering-based sampling approach for refreshing search engine's
  database.
\newblock In {\em Proceedings of the 10th International Workshop on the Web and
  Databases}, 2007.

\end{thebibliography}

\end{document}